\PassOptionsToPackage{unicode}{hyperref}
\PassOptionsToPackage{hyphens}{url}
\PassOptionsToPackage{dvipsnames,svgnames,x11names}{xcolor}
\documentclass[
]{article}
\usepackage{amsmath,amssymb}
\usepackage{lmodern}
\usepackage{iftex}
\ifPDFTeX
  \usepackage[T1]{fontenc}
  \usepackage[utf8]{inputenc}
  \usepackage{textcomp} 
\else 
  \usepackage{unicode-math}
  \defaultfontfeatures{Scale=MatchLowercase}
  \defaultfontfeatures[\rmfamily]{Ligatures=TeX,Scale=1}
\fi
\IfFileExists{upquote.sty}{\usepackage{upquote}}{}
\IfFileExists{microtype.sty}{
  \usepackage[]{microtype}
  \UseMicrotypeSet[protrusion]{basicmath} 
}{}
\makeatletter
\@ifundefined{KOMAClassName}{
  \IfFileExists{parskip.sty}{%
    \usepackage{parskip}
  }{
    \setlength{\parindent}{0pt}
    \setlength{\parskip}{6pt plus 2pt minus 1pt}}
}{
  \KOMAoptions{parskip=half}}
\makeatother
\usepackage{xcolor}
\NewDocumentCommand\citeproctext{}{}
\NewDocumentCommand\citeproc{mm}{%
  \begingroup\def\citeproctext{#2}\cite{#1}\endgroup}
\makeatletter
 \let\@cite@ofmt\@firstofone
 \def\@biblabel#1{}
 \def\@cite#1#2{{#1\if@tempswa , #2\fi}}
\makeatother
\newlength{\cslhangindent}
\newlength{\csllabelwidth}
\newenvironment{CSLReferences}[2] 
 {\begin{list}{}{%
  \setlength{\itemindent}{0pt}
  \setlength{\leftmargin}{0pt}
  \setlength{\parsep}{0pt}
  \ifodd #1
   \setlength{\leftmargin}{\cslhangindent}
   \setlength{\itemindent}{-1\cslhangindent}
  \fi
  \setlength{\itemsep}{#2\baselineskip}}}
 {\end{list}}
\usepackage{calc}

\ifLuaTeX
\usepackage[bidi=basic]{babel}
\else
\usepackage[bidi=default]{babel}
\fi
\babelprovide[main,import]{american}

\def\languageshorthands#1{}
\ifLuaTeX
  \usepackage{selnolig}  
\fi
\IfFileExists{bookmark.sty}{\usepackage{bookmark}}{\usepackage{hyperref}}
\IfFileExists{xurl.sty}{\usepackage{xurl}}{} 
\hypersetup{
  pdftitle={spherical: A Comprehensive Database and Automated Pipeline
for VLT/SPHERE High-Contrast Imaging},
  pdfauthor={Matthias Samland},
  pdflang={en-US},
  colorlinks=true,
  linkcolor={Maroon},
  filecolor={Maroon},
  citecolor={Blue},
  urlcolor={Blue},
  pdfcreator={LaTeX via pandoc}}

\title{spherical: A Comprehensive Database and Automated Pipeline for
VLT/SPHERE High-Contrast Imaging}

\definecolor{c53baa1}{RGB}{83,186,161}
\definecolor{c202826}{RGB}{32,40,38}

\usepackage[affil-it]{authblk}
\usepackage{orcidlink}
\author[1%
  ]{Matthias Samland%
    \,\orcidlink{0000-0001-9992-4067}\,%
    }

\affil[1]{Max-Planck-Institut für Astronomie (MPIA), Königstuhl 17,
69117 Heidelberg, Germany%
  }
\date{11 August 2026}

\begin{document}
\maketitle

\section{Summary}\label{summary}

The Spectro-Polarimetric High-contrast Exoplanet REsearch instrument
(SPHERE; Beuzit et al. (\citeproc{ref-Beuzit:2019}{2019})) at the Very
Large Telescope (VLT) is a leading facility for coronagraphic imaging of
exoplanets and circumstellar disks in the optical and near-infrared.
Over the last decade, SPHERE has contributed to hundreds of publications
and major legacy surveys (e.g., the SpHere INfrared survey for
Exoplanets, SHINE; Chauvin et al. (\citeproc{ref-Chauvin:2017}{2017});
Chomez et al. (\citeproc{ref-Chomez:2025}{2025})).

\texttt{spherical} combines a curated, searchable database of the
complete SPHERE observation history with automated reduction pipelines
for two of its subsystems. Users can filter observations by target star
properties, observing mode, or observing conditions, download the
selected raw data and calibrations from the European Southern
Observatory (ESO) archive, and reduce them end to end. Integral Field
Spectrograph (IFS; Claudi et al. (\citeproc{ref-Claudi:2008}{2008}))
data are extracted into spectral cubes by a pipeline adapted from the
Coronagraphic High Angular Resolution Imaging Spectrograph (CHARIS;
Brandt et al. (\citeproc{ref-Brandt:2017}{2017}); Samland et al.
(\citeproc{ref-Samland:2022}{2022})); data from the Infra-Red Dual-band
Imager and Spectrograph (IRDIS; Dohlen et al.
(\citeproc{ref-Dohlen:2008}{2008})) in dual-band imaging (DBI) mode are
calibrated and pre-processed by a pipeline that mirrors the IFS
workflow. Both feed post-processing with TRAP (Temporal Reference
Analysis of Planets; Samland et al.
(\citeproc{ref-Samland:2021}{2021})), a detection algorithm that
identifies companions by modeling the temporal systematics of the
starlight instead of subtracting a reference image, and recovers their
astrometry and contrast spectra.

As of the August 2026 database release, the database covers observations
taken between 2014 May and 2026 August: 6101 IRDIS DBI sequences, 4733
IFS sequences, 1205 IRDIS dual-beam polarimetric imaging (DPI; de Boer
et al. (\citeproc{ref-deBoer:2020}{2020}); van Holstein et al.
(\citeproc{ref-vanHolstein:2020}{2020a})) sequences, and 355 sparse
aperture masking (SAM; Cheetham et al.
(\citeproc{ref-Cheetham:2016}{2016})) sequences across both subsystems.
DPI and SAM data are covered for discovery and download only; their
reduction is left to community tools. SPHERE is undergoing the
\emph{SPHERE+} upgrade (\citeproc{ref-Boccaletti:2022}{Boccaletti et
al., 2022}, \citeproc{ref-Boccaletti:2020}{2020}), including a
second-stage adaptive optics system (SAXO+; Stadler et al.
(\citeproc{ref-Stadler:2022}{2022})), which extends its scientific
lifetime and makes it a pathfinder for the Planetary Camera and
Spectrograph (PCS; Kasper et al. (\citeproc{ref-Kasper:2021}{2021})) of
the Extremely Large Telescope (ELT).

\section{Statement of Need}\label{statement-of-need}

The ESO VLT/SPHERE archive is the world's largest collection of
high-contrast imaging data for detecting exoplanets, substellar
companions, and circumstellar disks. Using it end to end remains
difficult. The observation history is not available in a form that can
be queried by scientific criteria, only by header keywords such as
target name or coordinates. Raw data and their calibrations must be
queried and retrieved by hand. These barriers are most challenging for
researchers assembling large homogeneous samples for population studies
or for uniform extraction of exoplanet spectra, and for survey teams
needing rapid follow-up. \texttt{spherical} is written for observers
working on direct imaging of exoplanets and disks, SPHERE survey teams,
and archival researchers.

\section{State of the Field}\label{state-of-the-field}

Several existing tools address isolated components of the SPHERE data
workflow. The \textbf{High Contrast Data Center (DC)}
(\citeproc{ref-Delorme:2017}{Delorme et al., 2017}) reduces datasets
with the ESO pipeline recipes behind a Java-based interface, but offers
limited capabilities for programmatic interaction or custom batch
processing. \textbf{vlt-sphere} (\citeproc{ref-Vigan:2020}{Vigan, 2020})
drives the same recipes through \texttt{esorex} for user-provided raw
data, and therefore requires a local ESO pipeline installation. It lacks
automated archival download and integrated post-processing. For
polarimetry, \textbf{IRDAP} (\citeproc{ref-Holstein:2020ascl}{van
Holstein et al., 2020b}) is a well-established automated pipeline,
though it requires the user to discover and retrieve the raw data
manually.

What none of them provides is a high-level view of what SPHERE has
observed over its lifetime. \texttt{spherical} supplies that view and
connects it to a single workflow covering identification, download,
reduction, and post-processing with TRAP. From selecting observations of
interest through to detecting companions and extracting their spectra,
it is an end-to-end framework. It can equally serve as the retrieval and
pre-processing stage for alternative post-processing ecosystems such as
VIP (\citeproc{ref-Christiaens:2023}{Christiaens et al., 2023};
\citeproc{ref-Gonzales:2017}{Gomez Gonzalez et al., 2017}), pyKLIP
(\citeproc{ref-Wang:2015ascl}{Wang et al., 2015}), and IRDAP.

\section{Software Design}\label{software-design}

The architecture of \texttt{spherical} is a high-level abstraction layer
over the ESO raw archive, providing a systematic interface that does not
currently exist for high-contrast imaging data. A central design choice
was to decouple metadata curation from data reduction. Science headers
ingested from the ESO archive are cross-matched on telescope pointing
against SIMBAD (\citeproc{ref-Wenger:2000}{Wenger et al., 2000}) and
Gaia DR3 (\citeproc{ref-GaiaDR3:2023}{Gaia Collaboration et al., 2023}).
Because archival metadata is noisy, target ambiguities are resolved by
proximity, brightness, and object type, with stellar proper motion
propagated to the epoch of observation.

The result is a structured, searchable observation table, available
pre-computed on Zenodo
(\href{https://doi.org/10.5281/zenodo.15147730}{DOI:
10.5281/zenodo.15147730}). It aggregates observing mode, total exposure
time, parallactic angle coverage, and atmospheric conditions.
Accompanying target tables carry Gaia DR3 astrophysical parameters and,
via a cross-match against MOCAdb (\citeproc{ref-Gagne:2026}{Gagné et
al., 2026}), young-association membership, BANYAN \(\Sigma\) membership
probabilities, and adopted ages, so samples can be selected on stellar
age and association rather than on pointing, conditions, and stellar
parameters alone. Choosing a local, curated table over live archive
queries makes complex filtering of the entire instrument history fast.
The tables can be built from scratch, but the recommended entry point is
the published set, which a single command extends with data taken since
the last release. Every build records its provenance, both in a separate
file and in FITS header keywords. The record includes the
\texttt{spherical} version, the ESO query date and coverage range, the
Gaia data release, the status of each enrichment source, and the build
parameters. A published table set is therefore self-describing, and a
result can be traced back to the exact database state that produced it.

Rather than reimplementing reduction algorithms, \texttt{spherical}
wraps them, providing a Pythonic interface to the IFS cube extraction of
Samland et al. (\citeproc{ref-Samland:2022}{2022}) and to TRAP. This
approach prioritizes scientific continuity and maintainability: gluing
both into one object-oriented framework turns specialized code into a
workflow that can be run over hundreds of datasets, without altering the
algorithms or their published behavior. The reduction does not depend on
ESO's pipeline software. Calibration, extraction, and pre-processing are
implemented in Python, so the full chain installs with \texttt{pip}
without \texttt{esorex} or ESO pipeline kits, and every step can be
inspected and modified in one language.

The IRDIS dual-band pipeline introduced in v3.0.0 mirrors the IFS one,
i.e., per-observation master calibrations, pre-processing into
calibrated cubes with analytic inverse-variance maps, centering from the
satellite-spot (waffle) frames with offsets propagated through the
dithering header keywords for dithered sequences, photometric
calibration from the unocculted point spread function (PSF), and TRAP
post-processing with spectral template matching across the two channels.
Both subsystems share one configuration model. They resume by default,
so an interrupted campaign continues where it stopped. The IRDIS
reduction reproduces the published photometry and astrometry of 51
Eridani b, and the IFS reduction recovers a consistent position from the
simultaneous IFS data. Both results are pinned by regression tests
against frozen baselines.

\texttt{spherical} relies only on community Python packages:
\texttt{Astropy} (\citeproc{ref-Astropy:2013}{Astropy Collaboration et
al., 2013}, \citeproc{ref-Astropy:2018}{2018},
\citeproc{ref-Astropy:2022}{2022}), \texttt{astroquery} for ESO archive
and catalog access (\citeproc{ref-Ginsburg:2019}{Ginsburg et al.,
2019}), \texttt{NumPy} (\citeproc{ref-Harris:2020}{Harris et al., 2020})
for numerical operations, and \texttt{pandas} (\citeproc{ref-Pandas}{The
pandas development team, 2020}) for tabular data handling. The IFS
reduction uses the adapted \texttt{CHARIS} pipeline
(\citeproc{ref-Brandt:2017}{Brandt et al., 2017};
\citeproc{ref-Samland:2022}{Samland et al., 2022}), with calibration
routines derived in part from Vigan (\citeproc{ref-Vigan:2020}{2020})
(see the \texttt{vlt-sphere} repository for individual contributors).
Stellar effective temperatures for TRAP's spectral templates come from
Gaia DR3 where available and otherwise from a tabulation of Pecaut \&
Mamajek (\citeproc{ref-Pecaut:2013}{2013}). Post-processing employs
\texttt{TRAP} (\citeproc{ref-Samland:2021}{Samland et al., 2021}).

\section{Research Impact Statement}\label{research-impact-statement}

\texttt{spherical} has been used in several published peer-reviewed
studies to analyze SPHERE IFS data (\citeproc{ref-Franson:2023}{Franson
et al., 2023}; \citeproc{ref-Hammond:2025}{Hammond et al., 2025};
\citeproc{ref-Stolker:2025}{Stolker et al., 2025}), improving the
extraction of spectra of exoplanetary atmospheres. At the time these
works were published, the code was not yet citable via an existing DOI.
Since the database of existing observations was made available on
Zenodo, it has been downloaded more than 400 times. New exoplanet
candidates identified through \texttt{spherical} have been accepted for
follow-up observation in an ESO P117 proposal (117.2A06.001), awarding
28.5 hours, more than 60\% of all time awarded in P117 to SPHERE
observations.

Reducing the effort needed to go from raw archival data to science-ready
products makes it practical to assemble the homogeneous samples that
exoplanet occurrence-rate and population studies require, and to
reprocess the full twelve-year archive with one algorithm in search of
companions missed by earlier, less sensitive techniques. Analyses are
reusable and reproducible by others. Database releases are versioned and
record their provenance, so an analysis can name the exact database
state it used and a reader can retrieve it. The reductions are checked
against 51 Eridani b, and the comparisons are documented alongside the
regression tests.

\section{Future Work}\label{future-work}

Database coverage and retrieval will be extended to the remaining
observing modes, the Zurich Imaging Polarimeter (ZIMPOL; Schmid et al.
(\citeproc{ref-Schmid:2018}{2018})) and IRDIS long-slit spectroscopy
(\citeproc{ref-Vigan:2008}{Vigan et al., 2008}), with the respective
community reduction pipelines integrated into the same script-driven
framework. Archive-facing tests already replay recorded queries and the
51 Eridani baselines cover post-processing; a public raw test dataset
would extend continuous integration to the reduction stages themselves.
The package will also track the SPHERE+ upgrade, adapting calibration
and post-processing to the instrument's improved performance and to
SAXO+.

\section{AI Usage Disclosure}\label{ai-usage-disclosure}

Generative AI was used in the development of \texttt{spherical} and in
the preparation of this manuscript. The system used was Claude Code
(Anthropic), with Claude Sonnet and Claude Opus models of the 4.x and 5
model families, over the period from mid-2025 to the v3.1.0 release in
2026.

Within the software, AI assistance was concentrated in the
infrastructure surrounding the scientific algorithms rather than in the
algorithms themselves: pipeline orchestration and step sequencing, the
structured logging and crash-report system, the typed configuration
dataclasses, the console entry points and the database build, update,
and provenance workflow, the test suite, and docstrings, README, and
changelog text. In preparing this manuscript, generative AI was used for
language editing and to improve conciseness, not to draft its scientific
content.

AI was not used for the scientific methodology. The spectral extraction
(\citeproc{ref-Samland:2022}{Samland et al., 2022}) and post-processing
(\citeproc{ref-Samland:2021}{Samland et al., 2021}) algorithms are
pre-existing published work and are wrapped unchanged. The calibration
methodology, the target-resolution logic that disambiguates archival
pointings, the database schema, the decision to wrap rather than
reimplement, and the validation strategy are the author's own.

The author reviewed, edited, and validated all AI-assisted output, and
made the primary architectural and scientific design decisions.
Correctness was established scientifically rather than by code
inspection alone: the IRDIS pipeline reproduces the published photometry
and astrometry of 51 Eridani b, and both pipelines are covered by
regression tests pinned to frozen baselines.

\subsection{Acknowledgements}\label{acknowledgements}

I thank Lukas Welzel for motivating the public release, Elisabeth
Matthews for beta testing, and Tomas Stolker for testing and
contributing bug fixes. Contributors are listed at the project
repository's
\href{https://github.com/m-samland/spherical/graphs/contributors}{contributors
page}. We acknowledge ESO for SPHERE datasets and thank the developers
of \texttt{CHARIS}, \texttt{TRAP}, \texttt{Astropy}, and
\texttt{astroquery}.

\subsection*{References}\label{references}
\addcontentsline{toc}{subsection}{References}

\protect\phantomsection\label{refs}


\begin{CSLReferences}{1}{0}
\bibitem[\citeproctext]{ref-Astropy:2022}
Astropy Collaboration, Price-Whelan, A. M., Lim, P. L., Earl, N.,
Starkman, N., Bradley, L., Shupe, D. L., Patil, A. A., Corrales, L.,
Brasseur, C. E., Nöthe, M., Donath, A., Tollerud, E., Morris, B. M.,
Ginsburg, A., Vaher, E., Weaver, B. A., Tocknell, J., Jamieson, W.,
\ldots{} Astropy Project Contributors. (2022). {The Astropy Project:
Sustaining and Growing a Community-oriented Open-source Project and the
Latest Major Release (v5.0) of the Core Package}. \emph{The
Astrophysical Journal}, \emph{935}(2), 167.
\url{https://doi.org/10.3847/1538-4357/ac7c74}

\bibitem[\citeproctext]{ref-Astropy:2018}
Astropy Collaboration, Price-Whelan, A. M., Sipőcz, B. M., Günther, H.
M., Lim, P. L., Crawford, S. M., Conseil, S., Shupe, D. L., Craig, M.
W., Dencheva, N., Ginsburg, A., VanderPlas, J. T., Bradley, L. D.,
Pérez-Suárez, D., de Val-Borro, M., Aldcroft, T. L., Cruz, K. L.,
Robitaille, T. P., Tollerud, E. J., \ldots{} Astropy Contributors.
(2018). {The Astropy Project: Building an Open-science Project and
Status of the v2.0 Core Package}. \emph{The Astronomical Journal},
\emph{156}(3), 123. \url{https://doi.org/10.3847/1538-3881/aabc4f}

\bibitem[\citeproctext]{ref-Astropy:2013}
Astropy Collaboration, Robitaille, T. P., Tollerud, E. J., Greenfield,
P., Droettboom, M., Bray, E., Aldcroft, T., Davis, M., Ginsburg, A.,
Price-Whelan, A. M., Kerzendorf, W. E., Conley, A., Crighton, N.,
Barbary, K., Muna, D., Ferguson, H., Grollier, F., Parikh, M. M., Nair,
P. H., \ldots{} Streicher, O. (2013). {Astropy: A community Python
package for astronomy}. \emph{Astronomy \& Astrophysics}, \emph{558},
A33. \url{https://doi.org/10.1051/0004-6361/201322068}

\bibitem[\citeproctext]{ref-Beuzit:2019}
Beuzit, J.-L., Vigan, A., Mouillet, D., Dohlen, K., Gratton, R.,
Boccaletti, A., Sauvage, J.-F., Schmid, H. M., Langlois, M., Petit, C.,
Baruffolo, A., Feldt, M., Milli, J., Wahhaj, Z., Abe, L., Anselmi, U.,
Antichi, J., Barette, R., Baudrand, J., \ldots{} Zurlo, A. (2019).
{SPHERE: the exoplanet imager for the Very Large Telescope}.
\emph{Astronomy \& Astrophysics}, \emph{631}, A155.
\url{https://doi.org/10.1051/0004-6361/201935251}

\bibitem[\citeproctext]{ref-Boccaletti:2020}
Boccaletti, A., Chauvin, G., Mouillet, D., Absil, O., Allard, F.,
Antoniucci, S., Augereau, J.-C., Barge, P., Baruffolo, A., Baudino,
J.-L., Baudoz, P., Beaulieu, M., Benisty, M., Beuzit, J.-L., Bianco, A.,
Biller, B., Bonavita, B., Bonnefoy, M., Bos, S., \ldots{} Zurlo, A.
(2020). {SPHERE+: Imaging young Jupiters down to the snowline}.
\emph{arXiv e-Prints}, arXiv:2003.05714.
\url{https://doi.org/10.48550/arXiv.2003.05714}

\bibitem[\citeproctext]{ref-Boccaletti:2022}
Boccaletti, A., Chauvin, G., Wildi, F., Milli, J., Stadler, E.,
Diolaiti, E., Gratton, R., Vidal, F., Loupias, M., Langlois, M.,
Cantalloube, F., N'Diaye, M., Gratadour, D., Ferreira, F., Tallon, M.,
Mazoyer, J., Segransan, D., Mouillet, D., Beuzit, J.-L., \ldots{}
Zanutta, A. (2022). {Upgrading the high contrast imaging facility
SPHERE: science drivers and instrument choices}. In C. J. Evans, J. J.
Bryant, \& K. Motohara (Eds.), \emph{Ground-based and airborne
instrumentation for astronomy IX} (Vol. 12184, p. 121841S).
\url{https://doi.org/10.1117/12.2630154}

\bibitem[\citeproctext]{ref-Brandt:2017}
Brandt, T. D., Rizzo, M., Groff, T., Chilcote, J., Greco, J. P., Kasdin,
N. J., Limbach, M. A., Galvin, M., Loomis, C., Knapp, G., McElwain, M.
W., Jovanovic, N., Currie, T., Mede, K., Tamura, M., Takato, N., \&
Hayashi, M. (2017). {Data reduction pipeline for the CHARIS
integral-field spectrograph I: detector readout calibration and data
cube extraction}. \emph{Journal of Astronomical Telescopes, Instruments,
and Systems}, \emph{3}, 048002.
\url{https://doi.org/10.1117/1.JATIS.3.4.048002}

\bibitem[\citeproctext]{ref-Chauvin:2017}
Chauvin, G., Desidera, S., Lagrange, A.-M., Vigan, A., Feldt, M.,
Gratton, R., Langlois, M., Cheetham, A., Bonnefoy, M., \& Meyer, M.
(2017). {SHINE, The SpHere INfrared survey for Exoplanets}. In C. Reylé,
P. Di Matteo, F. Herpin, E. Lagadec, A. Lançon, Z. Meliani, \& F. Royer
(Eds.), \emph{SF2A-2017: Proceedings of the annual meeting of the french
society of astronomy and astrophysics} (p. 331).

\bibitem[\citeproctext]{ref-Cheetham:2016}
Cheetham, A. C., Girard, J., Lacour, S., Schworer, G., Haubois, X., \&
Beuzit, J.-L. (2016). {Sparse aperture masking with SPHERE}. In F.
Malbet, M. J. Creech-Eakman, \& P. G. Tuthill (Eds.), \emph{Optical and
infrared interferometry and imaging v} (Vol. 9907, p. 99072T).
\url{https://doi.org/10.1117/12.2231983}

\bibitem[\citeproctext]{ref-Chomez:2025}
Chomez, A., Delorme, P., Lagrange, A.-M., Gratton, R., Flasseur, O.,
Chauvin, G., Langlois, M., Mazoyer, J., Zurlo, A., Desidera, S., Mesa,
D., Bonnefoy, M., Feldt, M., Hagelberg, J., Meyer, M., Vigan, A.,
Ginski, C., Kenworthy, M., Albert, D., \ldots{} Wildi, F. (2025). {The
SPHERE infrared survey for exoplanets (SHINE): IV. Complete
observations, data reduction and analysis, detection performances, and
final results}. \emph{Astronomy \& Astrophysics}, \emph{697}, A99.
\url{https://doi.org/10.1051/0004-6361/202451751}

\bibitem[\citeproctext]{ref-Christiaens:2023}
Christiaens, V., Gonzalez, C., Farkas, R., Dahlqvist, C.-H., Nasedkin,
E., Milli, J., Absil, O., Ngo, H., Cantero, C., Rainot, A., Hammond, I.,
Bonse, M., Cantalloube, F., Vigan, A., Kompella, V., \& Hancock, P.
(2023). {VIP: A Python package for high-contrast imaging}. \emph{The
Journal of Open Source Software}, \emph{8}(81), 4774.
\url{https://doi.org/10.21105/joss.04774}

\bibitem[\citeproctext]{ref-Claudi:2008}
Claudi, R. U., Turatto, M., Gratton, R. G., Antichi, J., Bonavita, M.,
Bruno, P., Cascone, E., De Caprio, V., Desidera, S., Giro, E., Mesa, D.,
Scuderi, S., Dohlen, K., Beuzit, J. L., \& Puget, P. (2008). {SPHERE
IFS: the spectro differential imager of the VLT for exoplanets search}.
In I. S. McLean \& M. M. Casali (Eds.), \emph{Ground-based and airborne
instrumentation for astronomy II} (Vol. 7014, p. 70143E).
\url{https://doi.org/10.1117/12.788366}

\bibitem[\citeproctext]{ref-deBoer:2020}
de Boer, J., Langlois, M., van Holstein, R. G., Girard, J. H., Mouillet,
D., Vigan, A., Dohlen, K., Snik, F., Keller, C. U., Ginski, C., Stam, D.
M., Milli, J., Wahhaj, Z., Kasper, M., Schmid, H. M., Rabou, P., Gluck,
L., Hugot, E., Perret, D., \ldots{} Beuzit, J.-L. (2020). {Polarimetric
imaging mode of VLT/SPHERE/IRDIS. I. Description, data reduction, and
observing strategy}. \emph{Astronomy \& Astrophysics}, \emph{633}, A63.
\url{https://doi.org/10.1051/0004-6361/201834989}

\bibitem[\citeproctext]{ref-Delorme:2017}
Delorme, P., Meunier, N., Albert, D., Lagadec, E., Le Coroller, H.,
Galicher, R., Mouillet, D., Boccaletti, A., Mesa, D., Meunier, J.-C.,
Beuzit, J.-L., Lagrange, A.-M., Chauvin, G., Sapone, A., Langlois, M.,
Maire, A.-L., Montargès, M., Gratton, R., Vigan, A., \& Surace, C.
(2017). {The SPHERE Data Center: a reference for high contrast imaging
processing}. In C. Reylé, P. Di Matteo, F. Herpin, E. Lagadec, A.
Lançon, Z. Meliani, \& F. Royer (Eds.), \emph{SF2A-2017: Proceedings of
the annual meeting of the french society of astronomy and astrophysics}
(p. 347). \url{https://doi.org/10.48550/arXiv.1712.06948}

\bibitem[\citeproctext]{ref-Dohlen:2008}
Dohlen, K., Langlois, M., Saisse, M., Hill, L., Origne, A., Jacquet, M.,
Fabron, C., Blanc, J.-C., Llored, M., Carle, M., Moutou, C., Vigan, A.,
Boccaletti, A., Carbillet, M., Mouillet, D., \& Beuzit, J.-L. (2008).
{The infra-red dual imaging and spectrograph for SPHERE: design and
performance}. In I. S. McLean \& M. M. Casali (Eds.), \emph{Ground-based
and airborne instrumentation for astronomy II} (Vol. 7014, p. 70143L).
\url{https://doi.org/10.1117/12.789786}

\bibitem[\citeproctext]{ref-Franson:2023}
Franson, K., Bowler, B. P., Bonavita, M., Brandt, T. D., Chen, M.,
Samland, M., Zhang, Z., Lueber, A., Heng, K., Kitzmann, D., Wolf, T.,
Jones, B. A., Tran, Q. H., Bardalez Gagliuffi, D. C., Biller, B.,
Chilcote, J., Crepp, J. R., Dupuy, T. J., Faherty, J., \ldots{}
Theissen, C. A. (2023). {Astrometric Accelerations as Dynamical Beacons:
Discovery and Characterization of HIP 21152 B, the First T-dwarf
Companion in the Hyades}. \emph{The Astronomical Journal},
\emph{165}(2), 39. \url{https://doi.org/10.3847/1538-3881/aca408}

\bibitem[\citeproctext]{ref-Gagne:2026}
Gagné, J., Moranta, L., Faherty, J. K., Curtis, J. L., Bickle, T. P.,
Couture, D., Chiasson David, A., Christie, K., Lambier, S., Leclerc, E.,
Poliquin, L., Belzile, D., \& Mamajek, E. E. (2026). {The Montreal Open
Clusters and Associations (MOCA) Database: A Census of Nearby
Associations, Open Clusters, and Young Substellar Objects within 500 pc
of the Sun}. \emph{The Astrophysical Journal Supplement Series}.
\url{https://doi.org/10.48550/arXiv.2602.15695}

\bibitem[\citeproctext]{ref-GaiaDR3:2023}
Gaia Collaboration, Vallenari, A., Brown, A. G. A., Prusti, T., de
Bruijne, J. H. J., Arenou, F., Babusiaux, C., Biermann, M., Creevey, O.
L., Ducourant, C., Evans, D. W., Eyer, L., Guerra, R., Hutton, A.,
Jordi, C., Klioner, S. A., Lammers, U. L., Lindegren, L., Luri, X.,
\ldots{} et al. (2023). {Gaia Data Release 3. Summary of the content and
survey properties}. \emph{Astronomy \& Astrophysics}, \emph{674}, A1.
\url{https://doi.org/10.1051/0004-6361/202243940}

\bibitem[\citeproctext]{ref-Ginsburg:2019}
Ginsburg, A., Sipőcz, B. M., Brasseur, C. E., Cowperthwaite, P. S.,
Craig, M. W., Deil, C., Guillochon, J., Guzman, G., Liedtke, S., Lian
Lim, P., Lockhart, K. E., Mommert, M., Morris, B. M., Norman, H.,
Parikh, M., Persson, M. V., Robitaille, T. P., Segovia, J.-C., Singer,
L. P., \ldots{} a subset of astropy Collaboration. (2019). {astroquery:
An Astronomical Web-querying Package in Python}. \emph{The Astronomical
Journal}, \emph{157}(3), 98.
\url{https://doi.org/10.3847/1538-3881/aafc33}

\bibitem[\citeproctext]{ref-Gonzales:2017}
Gomez Gonzalez, C. A., Wertz, O., Absil, O., Christiaens, V., Defrère,
D., Mawet, D., Milli, J., Absil, P.-A., Van Droogenbroeck, M.,
Cantalloube, F., Hinz, P. M., Skemer, A. J., Karlsson, M., \& Surdej, J.
(2017). {VIP: Vortex Image Processing Package for High-contrast Direct
Imaging}. \emph{The Astronomical Journal}, \emph{154}(1), 7.
\url{https://doi.org/10.3847/1538-3881/aa73d7}

\bibitem[\citeproctext]{ref-Hammond:2025}
Hammond, I., Christiaens, V., Price, D. J., Blakely, D., Trevascus, D.,
Bonse, M. J., Cantalloube, F., Marleau, G.-D., Pinte, C., Juillard, S.,
Samland, M., Thompson, W., \& Wallace, A. (2025). {Keplerian motion of a
compact source orbiting the inner disc of PDS 70: a third protoplanet in
resonance with b and c?} \emph{Monthly Notices of the Royal Astronomical
Society}, \emph{539}(2), 1613--1627.
\url{https://doi.org/10.1093/mnras/staf586}

\bibitem[\citeproctext]{ref-Harris:2020}
Harris, C. R., Millman, K. J., Walt, S. J. van der, Gommers, R.,
Virtanen, P., Cournapeau, D., Wieser, E., Taylor, J., Berg, S., Smith,
N. J., Kern, R., Picus, M., Hoyer, S., Kerkwijk, M. H. van, Brett, M.,
Haldane, A., Río, J. F. del, Wiebe, M., Peterson, P., \ldots{} Oliphant,
T. E. (2020). Array programming with {NumPy}. \emph{Nature},
\emph{585}(7825), 357--362.
\url{https://doi.org/10.1038/s41586-020-2649-2}

\bibitem[\citeproctext]{ref-Kasper:2021}
Kasper, M., Cerpa Urra, N., Pathak, P., Bonse, M., Nousiainen, J.,
Engler, B., Heritier, C. T., Kammerer, J., Leveratto, S., Rajani, C.,
Bristow, P., Le Louarn, M., Madec, P.-Y., Ströbele, S., Verinaud, C.,
Glauser, A., Quanz, S. P., Helin, T., Keller, C., \ldots{} Raynaud,
H.-F. (2021). {PCS {\textemdash} A Roadmap for Exoearth Imaging with the
ELT}. \emph{The Messenger}, \emph{182}, 38--43.
\url{https://doi.org/10.18727/0722-6691/5221}

\bibitem[\citeproctext]{ref-Pecaut:2013}
Pecaut, M. J., \& Mamajek, E. E. (2013). {Intrinsic Colors,
Temperatures, and Bolometric Corrections of Pre-main-sequence Stars}.
\emph{The Astrophysical Journal Supplement Series}, \emph{208}(1), 9.
\url{https://doi.org/10.1088/0067-0049/208/1/9}

\bibitem[\citeproctext]{ref-Samland:2021}
Samland, M., Bouwman, J., Hogg, D. W., Brandner, W., Henning, T., \&
Janson, M. (2021). {TRAP: a temporal systematics model for improved
direct detection of exoplanets at small angular separations}.
\emph{Astronomy \& Astrophysics}, \emph{646}, A24.
\url{https://doi.org/10.1051/0004-6361/201937308}

\bibitem[\citeproctext]{ref-Samland:2022}
Samland, M., Brandt, T. D., Milli, J., Delorme, P., \& Vigan, A. (2022).
{Spectral cube extraction for the VLT/SPHERE IFS. Open-source pipeline
with full forward modeling and improved sensitivity}. \emph{Astronomy \&
Astrophysics}, \emph{668}, A84.
\url{https://doi.org/10.1051/0004-6361/202244587}

\bibitem[\citeproctext]{ref-Schmid:2018}
Schmid, H. M., Bazzon, A., Roelfsema, R., Mouillet, D., Milli, J.,
Menard, F., Gisler, D., Hunziker, S., Pragt, J., Dominik, C.,
Boccaletti, A., Ginski, C., Abe, L., Antoniucci, S., Avenhaus, H.,
Baruffolo, A., Baudoz, P., Beuzit, J. L., Carbillet, M., \ldots{} Wildi,
F. (2018). {SPHERE/ZIMPOL high resolution polarimetric imager. I. System
overview, PSF parameters, coronagraphy, and polarimetry}.
\emph{Astronomy \& Astrophysics}, \emph{619}, A9.
\url{https://doi.org/10.1051/0004-6361/201833620}

\bibitem[\citeproctext]{ref-Stadler:2022}
Stadler, E., Diolaiti, E., Schreiber, L., Cortecchia, F., Lombini, M.,
Loupias, M., Magnard, Y., De Rosa, A., Malaguti, G., Maurel, D.,
Morgante, G., Rabou, P., Rochat, S., Schiavone, F., Terenzi, L., Vidal,
F., Cantalloube, F., Gendron, E., Gratton, R., \ldots{} Boccaletti, A.
(2022). {SAXO+, a second-stage adaptive optics for SPHERE on VLT:
optical and mechanical design concept}. In L. Schreiber, D. Schmidt, \&
E. Vernet (Eds.), \emph{Adaptive optics systems VIII} (Vol. 12185, p.
121854E). \url{https://doi.org/10.1117/12.2629970}

\bibitem[\citeproctext]{ref-Stolker:2025}
Stolker, T., Samland, M., Waters, L. B. F. M., van den Ancker, M. E.,
Balmer, W. O., Lacour, S., Sitko, M. L., Wang, J. J., Nowak, M., Maire,
A.-L., Kammerer, J., Otten, G. P. P. L., Abuter, R., Amorim, A.,
Benisty, M., Berger, J.-P., Beust, H., Blunt, S., Boccaletti, A.,
\ldots{} Yazici, S. (2025). {Direct imaging discovery of a young giant
planet orbiting on Solar System scales}. \emph{Astronomy \&
Astrophysics}, \emph{700}, A21.
\url{https://doi.org/10.1051/0004-6361/202555064}

\bibitem[\citeproctext]{ref-Pandas}
The pandas development team. (2020). \emph{{pandas-dev/pandas: Pandas}}.
Zenodo. \url{https://doi.org/10.5281/zenodo.3509134}

\bibitem[\citeproctext]{ref-vanHolstein:2020}
van Holstein, R. G., Girard, J. H., de Boer, J., Snik, F., Milli, J.,
Stam, D. M., Ginski, C., Mouillet, D., Wahhaj, Z., Schmid, H. M.,
Keller, C. U., Langlois, M., Dohlen, K., Vigan, A., Pohl, A., Carbillet,
M., Fantinel, D., Maurel, D., Origné, A., \ldots{} Beuzit, J.-L.
(2020a). {Polarimetric imaging mode of VLT/SPHERE/IRDIS. II.
Characterization and correction of instrumental polarization effects}.
\emph{Astronomy \& Astrophysics}, \emph{633}, A64.
\url{https://doi.org/10.1051/0004-6361/201834996}

\bibitem[\citeproctext]{ref-Holstein:2020ascl}
van Holstein, R. G., Girard, J. H., de Boer, J., Snik, F., Milli, J.,
Stam, D. M., Ginski, C., Mouillet, D., Wahhaj, Z., Schmid, H. M.,
Keller, C. U., Langlois, M., Dohlen, K., Vigan, A., Pohl, A., Carbillet,
M., Fantinel, D., Maurel, D., Origné, A., \ldots{} Beuzit, J.-L.
(2020b). \emph{{IRDAP: SPHERE-IRDIS polarimetric data reduction
pipeline}}. Astrophysics Source Code Library, record ascl:2004.015.

\bibitem[\citeproctext]{ref-Vigan:2020}
Vigan, A. (2020). \emph{{vlt-sphere: Automatic VLT/SPHERE data reduction
and analysis}}. Astrophysics Source Code Library, record ascl:2009.002.
\url{https://doi.org/10.5281/zenodo.6563998}

\bibitem[\citeproctext]{ref-Vigan:2008}
Vigan, A., Langlois, M., Moutou, C., \& Dohlen, K. (2008). {Exoplanet
characterization with long slit spectroscopy}. \emph{Astronomy \&
Astrophysics}, \emph{489}(3), 1345--1354.
\url{https://doi.org/10.1051/0004-6361:200810090}

\bibitem[\citeproctext]{ref-Wang:2015ascl}
Wang, J. J., Ruffio, J.-B., De Rosa, R. J., Aguilar, J., Wolff, S. G.,
\& Pueyo, L. (2015). \emph{{pyKLIP: PSF Subtraction for Exoplanets and
Disks}}. Astrophysics Source Code Library, record ascl:1506.001.

\bibitem[\citeproctext]{ref-Wenger:2000}
Wenger, M., Ochsenbein, F., Egret, D., Dubois, P., Bonnarel, F., Borde,
S., Genova, F., Jasniewicz, G., Laloë, S., Lesteven, S., \& Monier, R.
(2000). {The SIMBAD astronomical database. The CDS reference database
for astronomical objects}. \emph{Astronomy \& Astrophysics Supplement
Series}, \emph{143}, 9--22. \url{https://doi.org/10.1051/aas:2000332}

\end{CSLReferences}
\end{document}